\documentclass[prb,amsfonts,amssymb,amsmath,showpacs]{revtex4}
\usepackage{graphicx}

\begin{document}

\title{Anyon exciton revisited: Exact solutions for a few-particle system}

\author{D.G.W.~Parfitt} \affiliation{School of Physics, University of
Exeter, Stocker Road, Exeter EX4 4QL, United Kingdom}

\author{M.E.~Portnoi}
\email[]{m.e.portnoi@ex.ac.uk}
\altaffiliation[Also at ]{A.F.~Ioffe
Physico-Technical Institute, St.~Petersburg, Russia.}
\affiliation{School of Physics, University of Exeter, Stocker Road,
Exeter EX4 4QL, United Kingdom}

\date{\today}


\begin{abstract}
The anyon exciton model is generalized to the case of a neutral 
exciton consisting of a valence hole and an arbitrary number
$N$ of fractionally-charged quasielectrons (anyons). 
A complete set of exciton basis functions is obtained and these 
functions are classified using a result from the theory of partitions.
Expressions are derived for the inter-particle interaction matrix 
elements of a six-particle system \mbox{$(N=5)$}, which describes an
exciton against the background of an incompressible quantum liquid 
with filling factor \mbox{$\nu=2/5$}.
Several exact results are obtained in a boson approximation,
including the binding energy of a \mbox{$(N+1)$}-particle exciton 
with zero in-plane momentum and zero internal angular momentum.
\end{abstract}

\maketitle

\section{Introduction}

The anyon exciton model (AEM), which considers a neutral exciton made up of 
a valence hole and several fractionally-charged quasielectrons with
fractional statistics (anyons), was proposed a decade ago by Rashba 
and Portnoi.\cite{Rashba1993}
It was later developed \cite{Portnoi1996} to model incompressible 
quantum liquids (IQLs) with filling factors \mbox{$\nu=1/3$} and $2/3$, 
and is valid at large separation between the hole and the 
two-dimensional electron gas (2DEG), so that the Coulomb field of 
the hole does not destroy the IQL.
The AEM was motivated by experiments on intrinsic photoluminescence (PL)
in the fractional quantum Hall regime 
\cite{Goldberg1988,Heiman1988,Turberfield1990} that
showed a pronounced double-peak structure in emission spectra, 
and it has provided major insights into the role of electron-hole 
separation in determining the optical spectra.
The model yields multiple-branch energy spectra and gives a full
classification of states for a four-particle system.
It also predicts an increase in ground-state angular momentum and 
flattening of the corresponding dispersion curve with increasing 
layer separation.
The AEM has also shown good agreement for intermediate electron-hole 
separation with numerical finite-size calculations by Apalkov \emph{et al.}
\cite{Apalkov1995b,Apalkov1995c}

At the time when the model was proposed, most experiments on intrinsic
PL were carried out at constant carrier density, with the filling
factor changed by varying the magnetic field.
However, the AEM is valid for fixed fractional filling factors only.
More recent experimental techniques enable the 2DEG density in the
PL experiments to be changed using applied gate voltages 
\cite{Plentz1996,Plentz1997,Yusa2001,Yusa2002} or by varying the 
light intensity.\cite{Ashkinadze2002}
Thus, it is now possible to manipulate the effective electron-hole
separation (in units of magnetic length) while keeping the filling
factor constant. 
So far, charged excitons have been observed for small electron-hole 
separation, but for large separations it is predicted that neutral
anyon excitons should be observed.

These recent developments have rekindled our interest in the theoretical
treatment of spatially-separated electron-hole systems.
In this paper we address some important mathematical features of the AEM.
In Sec.~\ref{S:one} we derive a general few-particle wavefunction
for a neutral exciton consisting of a valence hole and $N$ anyons.
In Sec.~\ref{S:two} we move to consider interaction matrix elements
for a six-particle anyon exciton, which should be useful for analysis
of intrinsic PL experiments at \mbox{$\nu=2/5$}. (Note that there has 
been some debate as to whether the ground state at \mbox{$\nu=1/5$} 
is actually an IQL - see, e.g., Ref.~\onlinecite{Ellis1992} and 
references therein.)
In the last Section, exact results for the general \mbox{$(N+1)$}-particle
case are obtained and discussed. It is shown that the exciton remains
bound even at large separations between the 2DEG and the hole.

\section{Exciton basis functions}\label{S:one}

\subsection{Preliminaries}

We consider an exciton consisting of a valence hole with charge $+e$
and $N$ anyons with charge $-e/N$ and  statistical factor $\alpha$.
The hole and anyons reside in two different layers, separated by a distance
of $h$ magnetic lengths, and are subject to a magnetic field 
\mbox{$\mathbf{H}=H\hat{\mathbf{z}}$} perpendicular to their planes of 
confinement.
Unless otherwise stated, we shall assume that the hole and the
quasielectrons are in their corresponding lowest Landau levels.

To simplify the description of the exciton, the hole and anyons may be
considered as moving in the same plane with coordinates $\mathbf{r}_h$
and $\mathbf{r}_j$, respectively, and the the layer separation $h$ can
be introduced later when considering the anyon-hole interaction. 

An exciton consisting of a hole and $N$ anyons, all in the lowest
Landau level, will have a total of \mbox{$N+1$} degrees of freedom. 
As the exciton is neutral, we can assign it an in-plane momentum 
$\mathbf{k}$, which absorbs two of these degrees of freedom.
For \mbox{$N\geqslant 2$} the exciton will have \mbox{$N-1$} 
\emph{internal} degrees of freedom, which results in internal quantum 
numbers and a multiple-branch energy spectrum. 

\subsection{Derivation of basis functions}

At this stage we consider the particles as non-interacting and introduce
interactions later as necessary.
We can therefore write the \mbox{$(N+1)$}-particle Hamiltonian as
\begin{equation}\label{hamil1}
\hat{H}_0=\frac{1}{2m_h}\left({\hat{\mathbf{p}}}_h-\frac{q_h
\mathbf{A}}{c}\right)^2+\sum_{j=1}^N\:\frac{1}{2m_a}\left(
{\hat{\mathbf{p}}}_j-\frac{q_a\mathbf{A}}{c}\right)^2,
\end{equation}
where $q_h=+e$ and $q_a=-e/N$ are the hole and anyon charges, respectively.

Choosing the symmetric gauge 
\begin{equation}
\mathbf{A}=\frac{1}{2}\left[\mathbf{H}\times\mathbf{r}\right],
\end{equation}
and scaling all distances with the magnetic length 
\mbox{$l_H=(c\hbar/eH)^{1/2}$}, we obtain
\begin{equation}\label{hamil3}
\hat{H}_0=\frac{1}{2m_h}\left(\frac{1}{i}\nabla_h-\left[\hat{\mathbf{z}}
\times\mathbf{r}_h\right]\right)^2+\sum_{j=1}^N\:\frac{1}{2m_a}
\left(\frac{1}{i}\nabla_j+\frac{1}{N}\left[\hat{\mathbf{z}}
\times\mathbf{r}_j\right]\right)^2,
\end{equation}
where, as usual, $e$, $\hbar$, $c$, and the dielectric constant
are assumed equal to unity.

\begin{figure}
\begin{center}
\includegraphics[width=15cm,keepaspectratio]{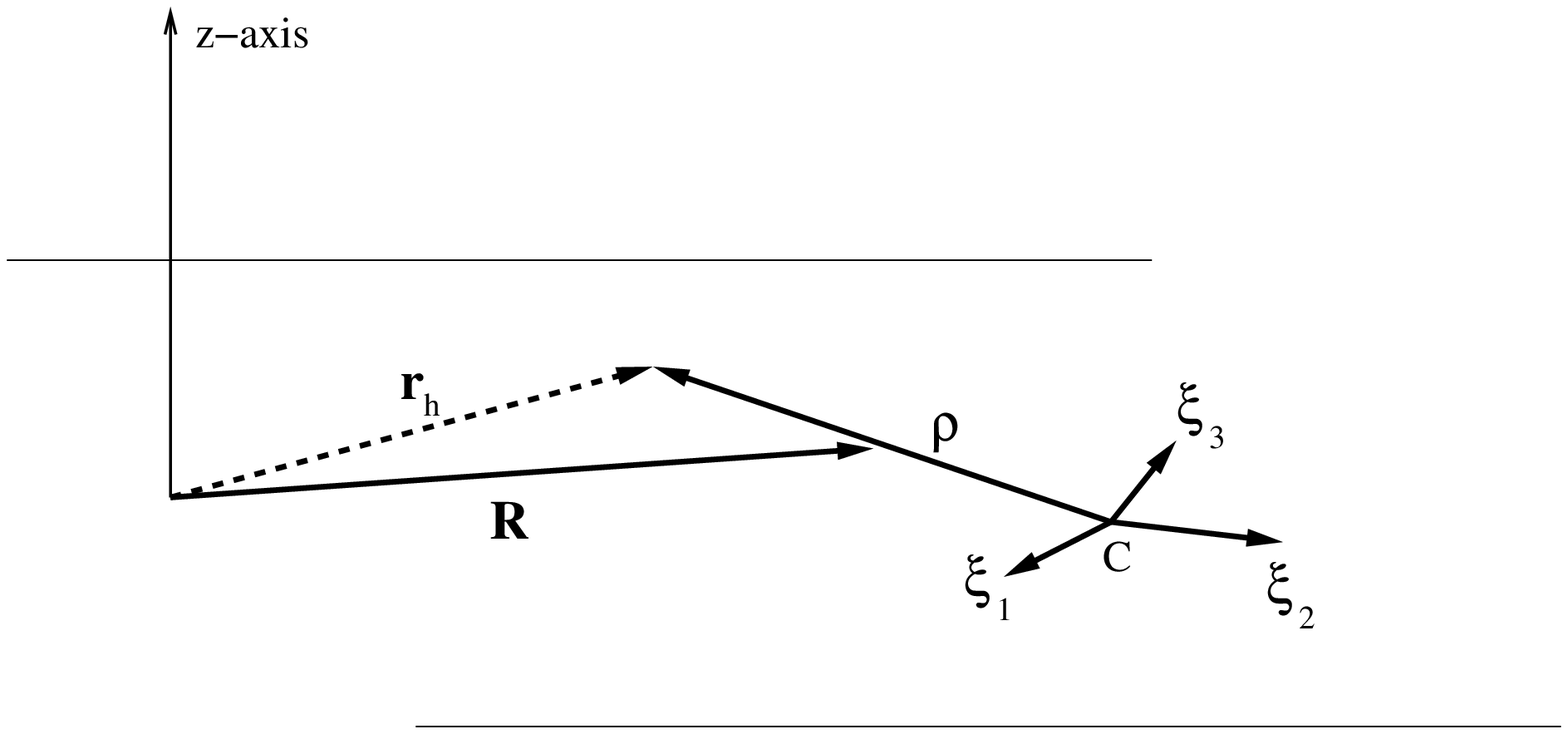}
\end{center}
\caption{Two-dimensional coordinate system for hole and several
quasielectrons (anyons). The hole is considered as being in the same plane 
as the anyons, with layer separation introduced when considering the
anyon-hole interaction. `C' indicates the centre of negative charge.}
\label{coords}
\vspace{1cm}
\end{figure}

We now introduce the following new coordinates (see Fig.~\ref{coords}):
\begin{equation}
\mathbf{R}=\frac{1}{2}\left(\mathbf{r}_h+\frac{1}{N}\sum_{j=1}^N\:
\mathbf{r}_j\right),\quad\boldsymbol{\rho}=\mathbf{r}_h-\frac{1}{N}
\sum_{j=1}^N\:\mathbf{r}_j,\quad\boldsymbol{\xi}_j=\mathbf{r}_j
-\frac{1}{N}\sum_{l=1}^N\:\mathbf{r}_l,
\end{equation}
together with the complex coordinates 
\mbox{$\zeta_j=\xi_{xj}+i\xi_{yj}$}.
Note also the following constraint on these coordinates:
\begin{equation}\label{constraint}
\sum_{j=1}^N\:\boldsymbol{\xi}_j=\sum_{j=1}^N\:\zeta_j=0.
\end{equation}

The fact that the variables $\boldsymbol{\xi}_j$ are not independent
means that derivatives with respect to these variables are not defined.
However, it is possible to introduce the derivatives 
$\nabla_{\boldsymbol{\xi}_j}$,
which treat the variables $\boldsymbol{\xi}_j$ as if they were
independent,\cite{Bolton1994} by working in a space of higher
dimensionality and then considering a submanifold in this space
defined by constraint \eqref{constraint}.

Using the chain rule, we may now rewrite the derivatives of a
function $f$ with respect to $\mathbf{r}_j$ as
\begin{align}\label{xideriv}
\nabla_j f&=\sum_{l=1}^N\:\nabla_{\boldsymbol{\xi}_l}
f(\ldots ,\boldsymbol{\xi}_l ,\ldots )\nabla_j\xi_l \notag \\
&=\nabla_{\boldsymbol{\xi}_j}f(\ldots ,\boldsymbol{\xi}_j ,\ldots )
-\frac{1}{N}\sum_{l=1}^N\:\nabla_{\boldsymbol{\xi}_l}f(\ldots ,
\boldsymbol{\xi}_l ,\ldots ).
\end{align}

The old coordinates can be expressed in terms of the new
ones as follows:
\begin{equation}\label{newvars2}
\mathbf{r}_h=\mathbf{R}+\frac{\boldsymbol{\rho}}{2},\quad
\mathbf{r}_j=\mathbf{R}-\frac{\boldsymbol{\rho}}{2}+
\boldsymbol{\xi}_j,
\end{equation}
and the corresponding derivatives as
\begin{align}
\nabla_h&=\frac{1}{2}\nabla_{\mathbf{R}}+
\nabla_{\boldsymbol{\rho}}, \\
\nabla_j&=\frac{1}{2N}\nabla_{\mathbf{R}}-\frac{1}{N}
\nabla_{\boldsymbol{\rho}}+\nabla_{\boldsymbol{\xi}_j}
-\frac{1}{N}\sum_{l=1}^N\:\nabla_{\boldsymbol{\xi}_l}.\label{nab2}
\end{align}
We now write the Hamiltonian \eqref{hamil3} in terms of the new 
variables.
Using Eqs.~\eqref{xideriv}\--\eqref{nab2} we obtain
\begin{equation}
\hat{H}_0=\hat{H}_{exc}+\hat{H}_{\xi},
\end{equation}
where
\begin{align}
\hat{H}_{exc}&=\frac{1}{2m_h}\left\{\frac{1}{2i}\nabla_{\mathbf{R}}
+\frac{1}{i}\nabla_{\boldsymbol{\rho}}-\frac{1}{2}\left[\hat{\mathbf{z}}
\times\left(\mathbf{R}+\frac{\boldsymbol{\rho}}{2}\right)\right]\right\}^2
\notag \\
&+\frac{1}{2Nm_a}\left\{\frac{1}{2i}\nabla_{\mathbf{R}}
-\frac{1}{i}\nabla_{\boldsymbol{\rho}}+\frac{1}{2}\left[\hat{\mathbf{z}}
\times\left(\mathbf{R}-\frac{\boldsymbol{\rho}}{2}\right)\right]\right\}^2,
\end{align}
and 
\begin{equation}\label{Hxi}
\hat{H}_{\xi}=\frac{1}{2m_a}\sum_{j=1}^N\left\{\frac{1}{i}
\nabla_{\boldsymbol{\xi}_j}
-\frac{1}{Ni}\sum_{l=1}^N\nabla_{\boldsymbol{\xi}_l}+\frac{1}{2N}
\left[\hat{\mathbf{z}}\times\boldsymbol{\xi}_j\right]\right\}^2.
\end{equation}
Note that Eq.~\eqref{Hxi} was obtained by applying constraint 
\eqref{constraint}.

The first part $\hat{H}_{exc}$ is similar to that for a standard
diamagnetic exciton,\cite{Gorkov1968,Lerner1980} with the electron mass
replaced with $N$ anyon masses. The eigenfunctions of this operator can be 
written straightforwardly as
\begin{equation}\label{wfpsi}
\Psi_{exc}(\mathbf{R} ,\boldsymbol{\rho})=\exp\left\{i\mathbf{k}
\cdot\mathbf{R}+i\hat{\mathbf{z}}\cdot\left[\mathbf{R}\times\boldsymbol{\rho}
\right]/2\right\}
\Phi(\boldsymbol{\rho}),
\end{equation}
where $\Phi(\boldsymbol{\rho})$ satisfies the equation
\begin{equation}
\left(\frac{1}{2m_h}\left\{\frac{1}{i}\nabla_{\boldsymbol{\rho}}
-\frac{\left[\hat{\mathbf{z}}
\times\left(\boldsymbol{\rho}-\mathbf{d}\right)\right]}{2}\right\}^2
+\frac{1}{2Nm_a}\left\{\frac{1}{i}\nabla_{\boldsymbol{\rho}}
+\frac{\left[\hat{\mathbf{z}}\times\left(\boldsymbol{\rho}-\mathbf{d}
\right)\right]}{2}\right\}^2\right)\Phi=E_{exc}\Phi.
\end{equation}
Here, $\mathbf{d}=\mathbf{k}\times\hat{\mathbf{z}}$ is the exciton 
dipole moment. Note that Eq.~\eqref{wfpsi} does not contain the 
mass-dependent phase factor that appears in the wavefunction in
Ref.~\onlinecite{Lerner1980}. This is due to our choice of the center of
charge of the exciton for our coordinate $\mathbf{R}$, rather than
the center of mass chosen in Refs.~\onlinecite{Gorkov1968,Lerner1980}.
Thus, the quantum number $\mathbf{k}$ entering Eq.~\eqref{wfpsi}
represents the momentum of the geometrical center of the exciton.

The eigenfunctions are then
\begin{equation}\label{eig1}
\Phi_{nm}=|\boldsymbol{\rho}-\mathbf{d}|^{|m|}L_n^{|m|}
\left\{(\boldsymbol{\rho}
-\mathbf{d})^2/2\right\}\exp\left\{im\phi-(\boldsymbol{\rho}
-\mathbf{d})^2/4\right\},
\end{equation}
where $\phi$ is the azimuthal angle of the vector 
\mbox{$(\boldsymbol{\rho}-\mathbf{d})$} and $L_n^m(z)$ are the 
associated Laguerre polynomials.
The corresponding eigenvalues are 
\begin{equation}\label{eig2}
E_{exc}=\frac{1}{m_h}\left(n+\frac{|m|-m+1}{2}\right)
+\frac{1}{Nm_a}\left(n+\frac{|m|+m+1}{2}\right).
\end{equation}
Note that in Eqs.~\eqref{eig1} and \eqref{eig2} we no longer
restricted ourselves to the lowest Landau level for the hole and
anyons.
For the ground-state \mbox{$(n,m=0)$}, Eq.~\eqref{wfpsi} reduces to
\begin{equation}
\Psi_{exc}(\mathbf{R} ,\boldsymbol{\rho})=\exp\left\{i\mathbf{k}\cdot
\mathbf{R}+i\hat{\mathbf{z}}\cdot\left[\mathbf{R}\times\boldsymbol{\rho}
\right]/2-(\boldsymbol{\rho}-
\mathbf{d})^2/4\right\},
\end{equation}
which depends on the quantum number $\mathbf{k}$ alone.

We now move to consider the term $\hat{H}_{\xi}$ in Eq.~\eqref{Hxi}, 
which may be divided into two parts:
\begin{equation}
\hat{H}_{1\xi}=\frac{1}{2m_{a}}\left(1-\frac{1}{N}\right)
\sum_{j=1}^N\left(\frac{1}{i}\nabla_{\boldsymbol{\xi}_j}+\frac{1}{2N}
\left[\hat{\mathbf{z}}\times{\boldsymbol{\xi}}_j\right]\right)^2,
\end{equation}
and
\begin{equation}
\hat{H}_{2\xi}=\frac{1}{2Nm_a}\sum_{j=1}^N\Bigg\{-
\nabla_{\boldsymbol{\xi}_j}^2+\frac{1}{Ni}\left[\hat{\mathbf{z}}\times
{\boldsymbol{\xi}}_j\right]\cdot\left(\nabla_{\boldsymbol{\xi}_j}-
\sum_{l=1}^N\nabla_{\boldsymbol{\xi}_l}\right)^{ }
+\frac{1}{N^2}|{\boldsymbol{\xi}_j}|^2
+2\nabla_{\boldsymbol{\xi}_j}\cdot
\left(\sum_{l=1}^N\nabla_{\boldsymbol{\xi}_l}\right)
-\frac{1}{N}\left(\sum_{l=1}^N\nabla_{\boldsymbol{\xi}_l}\right)^2
\Bigg\}.
\end{equation}

The eigenfunctions of the ground state of $\hat{H}_{1\xi}$ can then
be written in terms of complex coordinates as
\begin{equation}
\Phi_{\xi}=\prod_j\left(\bar{\zeta}_j\right)^{\beta_j}\exp\left\{
-|\zeta_j|^2/4N\right\},
\end{equation}
with the corresponding eigenvalues
\begin{equation}
E_{\xi}=\frac{1}{2m_{a}}\left(1-\frac{1}{N}\right).
\end{equation}
If we now apply constraint \eqref{constraint}, we find that the
function $\Phi_{\xi}$ is also an eigenfunction of $\hat{H}_{2\xi}$,
with a corresponding eigenvalue of zero, i.e.
\begin{equation}
\hat{H}_{2\xi}\Phi_{\xi}\propto\left(\sum_{j=1}^N
{\boldsymbol{\xi}}_j\right)^2\Phi_{\xi}=0.
\end{equation}
A calculation of the total energy of the ground state gives
\begin{equation}
E=E_{exc}+E_{\xi}=\frac{1}{2m_h}+\frac{1}{2m_{a}}=\frac{1}{2}(\omega_h
+N\omega_a),
\end{equation}
which is indeed the ground-state energy of the original 
\mbox{$(N+1)$}-particle Hamiltonian in Eq.~\eqref{hamil3}.

The most general form for the ground-state eigenfunction of $\hat{H}_0$
is
\vspace{3mm}
\begin{equation}
\Psi\left(\mathbf{R},\boldsymbol{\rho},\{\zeta_i\}\right)=
\exp\left\{i\mathbf{k}\cdot\mathbf{R}+i\hat{\mathbf{z}}\cdot
\left[\mathbf{R}\times\boldsymbol{\rho}
\right]/2-
(\boldsymbol{\rho}-\mathbf{d})^2/4\right\}
F(\ldots ,\bar{\zeta}_i,\ldots)\prod_p\exp\left\{
-|\zeta_p|^2/4N\right\},
\vspace{3mm}
\end{equation}
where $F$ must only be a function of the complex conjugates $\bar{\zeta}_i$.
We choose $F$ as follows to satisfy the interchange rules for anyons:
\begin{equation}
F(\ldots ,\bar{\zeta}_i,\ldots)=P_L(\ldots ,\bar{\zeta}_i,\ldots)
\prod_{j<l}(\bar{\zeta}_j-\bar{\zeta}_l)^\alpha,
\end{equation}
where $P_L$ is a symmetric polynomial of degree $L$ in the variables
$\bar{\zeta}_i$. 
Note that for \mbox{$\mathbf{k}=0$} the problem has rotational
symmetry about the $z$-axis, and the degree of the symmetric polynomial
$L$ is related to the exciton angular momentum 
[\mbox{$L_z=-L-N(N-1)\alpha/2$}].

We are now in a position to express the anyon exciton basis functions in
the final form
\vspace{3mm}
\begin{equation}\label{finalbasis}
\Psi\left(\mathbf{R},\boldsymbol{\rho},\{\zeta_i\}\right)=
\exp\left\{i\mathbf{k}\cdot\mathbf{R}+i\hat{\mathbf{z}}\cdot
\left[\mathbf{R}\times\boldsymbol{\rho}\right]/2-
(\boldsymbol{\rho}-\mathbf{d})^2/4\right\}
P_L(\ldots ,\bar{\zeta}_i,\ldots)
\prod_{j<l}(\bar{\zeta}_j-\bar{\zeta}_l)^\alpha\prod_p\exp\left\{
-|\zeta_p|^2/4N\right\}.
\end{equation}

\subsection{Symmetric polynomials}

We now consider the precise structure of the symmetric polynomials
$P_L$ which appear in Eq.~\eqref{finalbasis}.
To determine the symmetric basis polynomials of order $L$ we apply the 
\emph{fundamental theorem of symmetric polynomials},\cite{Birkhoff1977}
which states that any symmetric polynomial in $N$ variables, 
\mbox{$P(x_1,\ldots ,x_N)$}, can be uniquely expressed in terms of the 
elementary symmetric polynomials
\begin{equation}\label{elem}
\sigma_1=\sum_i x_i,\quad\sigma_2=\sum_{i<j}x_i x_j,\quad
\sigma_3=\sum_{i<j<k}x_i x_j x_k,
\quad\ldots\; ,\quad\sigma_N=x_1 x_2\cdots x_N.
\end{equation}
It is now apparent that all linearly-independent symmetric 
basis polynomials of a particular degree may be enumerated 
by considering the possible products of the elementary symmetric 
polynomials \eqref{elem}, as the total degree $L$ is the sum of the 
degrees of the constituent polynomials.
For example, the possible symmetric polynomials of degree four are 
constructed from $\sigma_1^4$, $\sigma_1^2\sigma_2$, $\sigma_1\sigma_3$, 
$\sigma_2^2$, and $\sigma_4$.

There are two further problems. The first is to determine the number
of possible products of elementary symmetric polynomials for
a total degree $L$. The second is to determine the structure of
these products.

To calculate the number of linearly-independent symmetric basis polynomials 
of degree $L$ which may be constructed from the elementary symmetric 
polynomials \mbox{$\sigma_1,\ldots ,\sigma_N$} we use a result from the 
theory of partitions.\cite{Slomson1991}

The number of ways of partitioning a number $L$ into parts of
size \mbox{$1,2,\ldots ,P$} is given by the coefficient of $x^L$ in the
expansion of
\begin{equation}\label{genprod}
\frac{1}{1-x}\cdot\frac{1}{1-x^2}\cdots\frac{1}{1-x^P}
=\prod_{k=1}^P\; \frac{1}{1-x^k},
\end{equation}
where each term in the product is known as a \emph{generating function}.
The number of ways of partitioning increases rapidly with $L$
and does not follow any pattern.
Furthermore, the different products of elementary symmetric
polynomials for a particular value of $L$ must be determined
by hand. An approximate formula for calculating the number of 
ways of partitioning a number does exist, the so-called
Hardy-Ramanujan formula,\cite{Hardy1918} but it is not applicable
to the present case as we have no polynomial of order one due
to constraint \eqref{constraint}. This constraint significantly
reduces the number of possible symmetric polynomials by removing
the first factor in the product \eqref{genprod}.
For example, for \mbox{$N=3$} the number of polynomials of degree $L$
is equal to the integer part of \mbox{$L/6+1$} for even $L$ and
the integer part of \mbox{$(L-3)/6+1$} for odd $L$, which corresponds 
to the result obtained in Ref.~\onlinecite{Portnoi1996}.

\subsection{Discussion}

The key difference between the current general formulation and that
outlined in Ref.~\onlinecite{Portnoi1996} for the four-particle case
\mbox{$(N=3)$} is the replacement of anyon difference coordinates 
$\bar{z}_{jl}$ by the new coordinates $\bar{\zeta}_i$. 
The principal advantage of the difference coordinates was that they
simplified the calculation of inter-anyon repulsion matrix elements,
as well as the form of the statistical factor in the exciton wavefunction.
However, they had the disadvantage that the classification of symmetric
polynomials was more difficult, as it was necessary to introduce a
Vandermonde determinant for odd-$L$ polynomials. For example, in the
simplest case \mbox{$L=3$}, even though the number of anyon coordinates 
is the same in both formulations, we have 
\mbox{$\sigma_3=(\bar{z}_{12}-\bar{z}_{23})(\bar{z}_{23}-\bar{z}_{31})
(\bar{z}_{31}-\bar{z}_{12})$}
in terms of difference coordinates, whereas we have a simple product 
\mbox{$\sigma_3=\bar{\zeta}_1\bar{\zeta}_2\bar{\zeta}_3$} in terms of the 
new coordinates.
For a number of anyons greater than three this disadvantage is
crucial, as the classification of polynomials in terms of $\bar{z}_{jl}$
becomes too cumbersome and the number of constraints on these 
coordinates is greater than one.

\section{Six-particle anyon exciton: Boson approximation}\label{S:two}

\subsection{Formulation}

We introduce an anyon exciton consisting of a hole and five 
anyons with charge $-e/5$. 
We also make a boson approximation so that the statistical factor 
\mbox{$\alpha=0$}. Our justification for this step is as follows.
It was shown in Ref.~\onlinecite{Portnoi1996} that for large values 
of $h$ (which is required for the AEM to be valid)
the statistical factor $\alpha$ becomes unimportant, and the
results for \mbox{$\alpha=0$} were very similar to those for 
\mbox{$\alpha=\pm 1/3$}.
We would therefore also expect this to be true for \mbox{$\alpha=\pm 1/N$}, 
as this is even closer to zero. From now on we consider a boson
approximation \mbox{$(\alpha=0)$} for anyon statistics.

The explicit form of the exciton basis functions is
\begin{equation}\label{basisfuncs}
\Psi_{L,M,\mathbf{k}}\left(\mathbf{R},\boldsymbol{\rho},\{\zeta_i\}
\right)=\exp\left\{i\mathbf{k}\cdot\mathbf{R}
+i\hat{\mathbf{z}}\cdot\left[\mathbf{R}\times\boldsymbol{\rho}
\right]/2-(\boldsymbol{\rho}-\mathbf{d})^2/4\right\}
P_{L,M}({\bar{\zeta}}_1,\ldots,{\bar{\zeta}}_5)\prod_{j=1}^5
\exp\left\{-|\zeta_j|^2/20\right\},
\end{equation}
where $M$ enumerates different linearly-independent symmetric polynomials
of degree $L$.

To construct the symmetric polynomials $P_{L,M}$ we need to consider only
the elementary symmetric polynomials $\sigma_2$, $\sigma_3$, $\sigma_4$,
and $\sigma_5$. Note that \mbox{$\sigma_1=0$} because of constraint
\eqref{constraint}. From the above, we find that the 
number of possible ways of constructing a polynomial of degree $L$ 
is therefore the coefficient of $x^L$ in the expansion
\begin{equation}
\frac{1}{1-x^2}\cdot\frac{1}{1-x^3}\cdot\frac{1}{1-x^4}\cdot\frac{1}{1-x^5}
=\prod_{k=2}^5\; \frac{1}{1-x^k}.
\end{equation}
The possible ways of constructing the first twelve polynomials are shown 
explicitly in Table~\ref{Table1}.
It can be seen that the number and complexity of the polynomials
increases rapidly with the degree $L$.
\begin{table}
\caption{Possible ways of constructing a symmetric polynomial $P_L$
from the elementary symmetric polynomials $\sigma_2$, $\sigma_3$, 
$\sigma_4$, and $\sigma_5$.}
\begin{center}
\begin{tabular}{|c|c|l|} \hline
Order, $L$ & No. of polynomials & Structure \\
\hline
0 & 1 & 1 \\
\hline
1 & 0 & - \\
\hline
2 & 1 & $\sigma_2$ \\
\hline
3 & 1 & $\sigma_3$ \\
\hline
4 & 2 & $\sigma_2^2$, $\sigma_4$ \\
\hline
5 & 2 & $\sigma_2\sigma_3$, $\sigma_5$ \\
\hline
6 & 3 & $\sigma_2^3$, $\sigma_2\sigma_4$, $\sigma_3^2$ \\
\hline
7 & 3 & $\sigma_2^2\sigma_3$, $\sigma_2\sigma_5$, $\sigma_3\sigma_4$ \\
\hline
8 & 5 & $\sigma_2^4$, $\sigma_2^2\sigma_4$, $\sigma_2\sigma_3^2$,
$\sigma_3\sigma_5$, $\sigma_4^2$ \\
\hline
9 & 5 & $\sigma_2^3\sigma_3$, $\sigma_2^2\sigma_5$,
$\sigma_2\sigma_3\sigma_4$, $\sigma_3^3$, $\sigma_4\sigma_5$ \\
\hline
10 & 7 & $\sigma_2^5$, $\sigma_2^3\sigma_4$, $\sigma_2^2\sigma_3^2$,
$\sigma_2\sigma_3\sigma_5$, $\sigma_2\sigma_4^2$, $\sigma_3^2\sigma_4$, 
$\sigma_5^2$ \\
\hline
11 & 7 & $\sigma_2^4\sigma_3$, $\sigma_2^3\sigma_5$, 
$\sigma_2^2\sigma_3\sigma_4$, $\sigma_2\sigma_3^3$,
$\sigma_2\sigma_4\sigma_5$, \\
\hspace{2cm} & \hspace{4cm} & $\sigma_3^2\sigma_5$, $\sigma_3\sigma_4^2$ \\
\hline
12 & 10 & $\sigma_2^6$, $\sigma_2^4\sigma_4$, $\sigma_2^3\sigma_3^2$, 
$\sigma_2^2\sigma_3\sigma_5$, $\sigma_2^2\sigma_4^2$,
$\sigma_2\sigma_3^2\sigma_4$, \\
\hspace{2cm} & \hspace{4cm} & $\sigma_2\sigma_5^2$, $\sigma_3^4$,
$\sigma_3\sigma_4\sigma_5$, $\sigma_4^3$ \\
\hline
\end{tabular}
\label{Table1}
\end{center}
\end{table}

It is evident from Eq.~\eqref{basisfuncs} that basis functions with 
different values of $\mathbf{k}$ are orthogonal.
We now expand a general exciton wavefunction for given $\mathbf{k}$ in 
terms of the complete set of basis functions \eqref{basisfuncs}:
\begin{equation}
\Phi=\sum_i\chi_i\Psi_i.
\end{equation}
We shall show in Sec.~\ref{SS:overlap} that functions with different 
$L$ are orthogonal. 
However, basis functions with the same value of $L$ but different $M$ 
are not necessarily orthogonal, so their scalar products 
\mbox{$\langle L,M|L,M^{\prime}\rangle$} will be non-zero.
We therefore write the Schr\"{o}dinger equation in matrix form as
\begin{equation}\label{mateqn}
\hat{H}\boldsymbol{\chi}=\varepsilon\hat{B}\boldsymbol{\chi},
\end{equation}
where $\hat{B}$ is the block-diagonal matrix of scalar products 
(the overlap matrix) and $\varepsilon$ is an energy eigenvalue.
The size of each block in $\hat{B}$ depends on the number of different
wavefunctions with given $L$. 
For example, in the case of a six-particle exciton the block 
corresponding to \mbox{$L=12$} will be of size \mbox{$10\times 10$}
(see Table~\ref{Table1}). 
Both the overlap matrix $\hat{B}$ and the Hamiltonian matrix $\hat{H}$ 
are diagonal in $\mathbf{k}$.
Note that for \mbox{$\mathbf{k}=0$} the matrix $\hat{H}$ takes the same
block-diagonal form as $\hat{B}$ (as will be shown in 
Secs.~\ref{SS:anyonanyon} and \ref{SS:anyonhole}), and as a result 
the problem becomes exactly soluble. 

We shall now proceed to evaluate the matrix elements in Eq.~\eqref{mateqn}.
As $\hat{H}$ and $\hat{B}$ are diagonal in $\mathbf{k}$, in what
follows we consider only the matrix elements diagonal in $\mathbf{k}$.
Since all terms in the polynomial $P_{L,M}$ are of the form of a product
of the coordinates ${\bar{\zeta}}_j$, we use the following functions
in monomials in the matrix element calculations:
\begin{equation}\label{mono}
\Psi_{\{n\},\mathbf{k}}\left(\mathbf{R},\boldsymbol{\rho},\{\zeta_i\}
\right)=C\exp\left\{i\mathbf{k}\cdot\mathbf{R}
+i\hat{\mathbf{z}}\cdot\left[\mathbf{R}\times\boldsymbol{\rho}
\right]/2-(\boldsymbol{\rho}-\mathbf{d})^2/4\right\}
{\bar{\zeta}}_1^{n_1}{\bar{\zeta}}_2^{n_2}
{\bar{\zeta}}_3^{n_3}{\bar{\zeta}}_4^{n_4}{\bar{\zeta}}_5^{n_5}
\prod_{j=1}^5\exp\left\{-|\zeta_j|^2/20\right\},
\end{equation}
where $\{n\}$ denotes the set of quantum numbers $n_1$ to $n_5$,
and the constant $C$ will be defined in Sec.~\ref{SS:overlap}.
The basis functions in terms of symmetric polynomials of degree $L$
can be obtained as a linear combination of the monomial functions 
\eqref{mono}, and \mbox{$n_1+n_2+n_3+n_4+n_5=L$}.

We also take into account constraint \eqref{constraint} via the
following transformation:
\begin{equation}
\delta\left(\boldsymbol{\xi}_1+\boldsymbol{\xi}_2+\boldsymbol{\xi}_3
+\boldsymbol{\xi}_4+\boldsymbol{\xi}_5\right)
=\int\frac{d\mathbf{f}}{(2\pi)^2}\,\exp\left\{i\mathbf{f}\cdot
\left(\boldsymbol{\xi}_1+\boldsymbol{\xi}_2+\boldsymbol{\xi}_3
+\boldsymbol{\xi}_4+\boldsymbol{\xi}_5\right)\right\}.
\end{equation}

\subsection{Overlap matrix elements}\label{SS:overlap}

To calculate the overlap matrix $\hat{B}$, let us consider
the scalar product of two monomial functions:
\vspace{3mm}
\begin{equation}
\langle\left\{n\right\}|\left\{n'\right\}\rangle=\int d\mathbf{R}
\int d\boldsymbol{\rho}\int\frac{d\mathbf{f}}{(2\pi)^2}
\int d\boldsymbol{\xi}_1 d\boldsymbol{\xi}_2 d\boldsymbol{\xi}_3 
d\boldsymbol{\xi}_4 d\boldsymbol{\xi}_5
\bar{\Psi}_{{\{n\},\mathbf{k}}}\Psi_{\{n'\},\mathbf{k}}
\exp\left\{i\mathbf{f}\cdot
\left(\boldsymbol{\xi}_1+\boldsymbol{\xi}_2+\boldsymbol{\xi}_3
+\boldsymbol{\xi}_4+\boldsymbol{\xi}_5\right)\right\}.
\vspace{3mm}
\end{equation}
The integration over $\mathbf{R}$ and $\boldsymbol{\rho}$ gives
a factor of $2\pi A$, 
and this leaves the following:
\begin{equation}
\langle\left\{n\right\}|\left\{n'\right\}\rangle=C^2(2\pi A)
\int\frac{d\mathbf{f}}{(2\pi)^2}\;\prod_{j=1}^5 M_{n_j n'_j}(\mathbf{f}),
\end{equation}
where
\begin{equation}
M_{m m^\prime}(\mathbf{f})=\int d\boldsymbol{\xi}\:
\zeta^m{\bar{\zeta}}^{m'}e^{-\xi^2/10+i\mathbf{f}\cdot
\boldsymbol{\xi}}.
\end{equation}
If we now make the substitution 
\mbox{$\zeta=\xi e^{i\phi_{\boldsymbol{\xi}}}$},
and let $\phi$ be the angle between $\mathbf{f}$ and $\boldsymbol{\xi}$, 
i.e.
\begin{equation}
\phi=\phi_{\boldsymbol{\xi}}-\phi_\mathbf{f},
\end{equation}
we obtain
\begin{equation}\label{phif}
M_{m m^\prime}(\mathbf{f})=e^{i(m-m')\phi_{\mathbf{f}}}
\int_0^{2\pi}d\phi\int_0^{\infty}d\xi\:\xi^{1+m+m'}
\exp\left\{i(m-m')\phi+if\xi\cos\phi-\xi^2/10\right\}.
\end{equation}
We can now make use of Bessel's integral:\cite{Arfken1985}
\begin{equation}
\int_0^{2\pi}d\phi\: e^{\pm i(m-m')\phi+if\xi\cos\phi}
=2\pi i^{|m-m'|}J_{|m-m'|}(f\xi),
\end{equation}
to reduce Eq.~\eqref{phif} to the form
\begin{equation}
M_{m m^\prime}(\mathbf{f})=2\pi i^{|m-m'|}e^{i(m-m')
\phi_{\mathbf{f}}}\mathcal{M}_{mm'}(f),
\end{equation}
where
\begin{equation}\label{curlym}
\mathcal{M}_{mm'}(f)=\int_0^{\infty}d\xi\:\xi^{1+m+m'}
e^{-\xi^2/10}J_{|m-m'|}(f\xi).
\end{equation}
This can also be expressed \cite{Gradshteyn2000} in terms of a 
confluent hypergeometric function \mbox{$\Phi(\beta,\gamma;z)$} as
\vspace{3mm}
\begin{equation}\label{confhyp}
\mathcal{M}_{mm'}(f)=\frac{2^{(m+m')/2}5^{(m+m')/2+1}
\Gamma\left(\max\{m,m'\}+1\right)}{|m-m'|!}t^{|m-m'|/2} 
\Phi\left(\max\{m,m'\}+1,|m-m'|+1;-t\right),
\vspace{3mm}
\end{equation}
where $t=5f^2/2$ and $\max\{m,m'\}$ is the largest of the integers
$m$ and $m'$. Eq.~\eqref{confhyp} can be simplified further if we apply 
the Kummer transformation
\begin{equation}
\Phi(\beta,\gamma;-t)=e^{-t}\Phi(\gamma-\beta,\gamma;t),
\end{equation} 
and note that \mbox{$(\gamma-\beta)$} is always a non-positive integer:
\begin{equation}
\gamma-\beta=|m-m'|-\max\{m,m'\}.
\end{equation}
This means that $\Phi(\gamma-\beta,\gamma;t)$ reduces to a polynomial,
and \mbox{$\Phi(\beta,\gamma;-t)$} is then just a polynomial in $t$ 
multiplied by $e^{-t}$.

If we now perform the integration over $\phi_{\mathbf{f}}$ we obtain the
final form of the overlap matrix elements
\begin{equation}\label{monom}
\langle\left\{n\right\}|\left\{n'\right\}\rangle=C^2A(2\pi)^5\;\delta_{LL'}
\int_0^{\infty}df\: f\prod_{j=1}^5 i^{|n_j-n'_j|}\mathcal{M}_{n_j n'_j}(f),
\end{equation}
where $L=n_1+n_2+n_3+n_4+n_5$.
Note that nothing depends on $\mathbf{k}$ in this expression.
It is convenient to define the constant $C$ so that for 
\mbox{$L=L^{\prime}=0$} the matrix element 
\mbox{$\langle\left\{n\right\}|\left\{n'\right\}\rangle=
\langle 0|0\rangle=1$}.
This yields
\begin{equation}\label{csquared}
C^2=\frac{1}{125(2\pi)^5A}.
\end{equation}

In the above formulation we have only considered matrix elements in
terms of monomial functions. As mentioned earlier, matrix elements in
terms of symmetric polynomials can be easily constructed as a linear 
combination of the monomial matrix elements \eqref{monom}.

\subsection{Anyon-anyon interaction}\label{SS:anyonanyon}

The anyon-anyon interaction has the form
\begin{equation}\label{Vaa}
\hat{V}_{aa}=\frac{1}{25}\sum_{j<l}\;\frac{1}{|\boldsymbol{\xi}_j
-\boldsymbol{\xi}_l|}.
\end{equation}
We shall only calculate the matrix elements for
the first term $\hat{V}_{12}$ as the others follow by analogy.
We begin by taking the Fourier transform of Eq.~\eqref{Vaa}:
\begin{equation}
\hat{V}_{12}\left({\boldsymbol{\xi}}_1,{\boldsymbol{\xi}}_2\right)=
\frac{1}{25}\int\frac{d\mathbf{q}}{2\pi q}
\exp\left\{i\mathbf{q}\cdot\left({\boldsymbol{\xi}}_1
-{\boldsymbol{\xi}}_2\right)\right\}.
\end{equation}
This gives a matrix element in terms of monomials
\begin{multline}
\langle\left\{n\right\}|\hat{V}_{12}|\left\{n'\right\}\rangle=
\frac{1}{25}\int d\mathbf{R}
\int d\boldsymbol{\rho}\int\frac{d\mathbf{f}}{(2\pi)^2}
\int\frac{d\mathbf{q}}{2\pi q}
\int d\boldsymbol{\xi}_1 d\boldsymbol{\xi}_2 d\boldsymbol{\xi}_3
d\boldsymbol{\xi}_4 d\boldsymbol{\xi}_5
\\ 
\times\bar{\Psi}_{{\{n\},\mathbf{k}}}\Psi_{\{n'\},\mathbf{k}}
\exp\left\{i\mathbf{f}\cdot
\left(\boldsymbol{\xi}_1+\boldsymbol{\xi}_2+\boldsymbol{\xi}_3
+\boldsymbol{\xi}_4+\boldsymbol{\xi}_5\right)\right\}
\exp\left\{i\mathbf{q}\cdot\left({\boldsymbol{\xi}}_1
-{\boldsymbol{\xi}}_2\right)\right\}.
\end{multline}
Following the procedure in Sec.~\ref{SS:overlap} yields
\begin{multline}
\langle\left\{n\right\}|\hat{V}_{12}|\left\{n'\right\}\rangle =
\frac{C^2A(2\pi)^3}{25}\prod_{j=1}^5 i^{|n_j-n'_j|}
\int d\mathbf{f}\int d\mathbf{q}\;\exp\left\{i\phi_{+}(n_1-n_1^{\prime})
+i\phi_{-}(n_2-n_2^{\prime})\right\} \\
\times\exp\left\{i\phi_{\mathbf{f}}\right[(n_3-n_3^{\prime})
+(n_4-n_4^{\prime})+(n_5-n_5^{\prime})\left]\right\}
\mathcal{M}_{n_1 n'_1}(f_{+})\mathcal{M}_{n_2 n'_2}(f_{-})
\prod_{j=3}^5\mathcal{M}_{n_j n'_j}(f),
\end{multline}
where $\mathcal{M}_{mm'}$ are as defined in Eq.~\eqref{curlym}, 
\mbox{$\mathbf{f}_{\pm}=\mathbf{f}\pm\mathbf{q}$}, and $\phi_{\pm}$
are the phases of $\mathbf{f}_{\pm}$.

We now seek to eliminate $\phi_{\mathbf{f}}$ by the change of variables
$\phi=\phi_{\mathbf{f}}-\phi_{\mathbf{q}}$ and $\psi_{\pm}=
\phi_{\pm}-\phi_{\mathbf{f}}$. After substitution and integration 
over $\phi_{\mathbf{q}}$ we obtain
\begin{multline}\label{aafinal}
\langle\left\{n\right\}|\hat{V}_{12}|\left\{n'\right\}\rangle =
\frac{C^2A(2\pi)^4}{25}\;\delta_{LL'}\prod_{j=1}^5 i^{|n_j-n'_j|}
\int_0^{\infty}df\: f\int_0^{\infty}dq\int_0^{2\pi}d\phi\;
\exp\left\{i\phi(L-L^{\prime})\right\} \\
\times
\exp\left\{i\psi_{+}(n_1-n_1^{\prime})+i\psi_{-}(n_2-n_2^{\prime})
\right\}
\mathcal{M}_{n_1 n'_1}(f_{+})\mathcal{M}_{n_2 n'_2}(f_{-})
\prod_{j=3}^5\mathcal{M}_{n_j n'_j},
\end{multline}
where $L=n_1+n_2+n_3+n_4+n_5$, and $\psi_{\pm}$ and $f_{\pm}$ can be 
expressed in terms of the variables of integration as
\begin{equation}
e^{i\psi_{\pm}}=\left(f\pm qe^{-i\phi}\right)/f_{\pm},\qquad
f_{\pm}^2=f^2+q^2\pm 2fq\cos\phi.
\end{equation}
It is evident from Eq.~\eqref{aafinal} that the anyon-anyon interaction
matrix has the same block-diagonal structure as the overlap matrix.
The integrand in Eq.~\eqref{aafinal} can be further reduced to a
product of a polynomial in $q$, $f$, and $e^{\pm i\phi}$, and an
exponential factor \mbox{$\exp(-25f^2/2-5q^2)$}.
This integral can therefore be evaluated analytically for any
$\{n\}$ and $\{n'\}$.
For the case of \mbox{$L=L^{\prime}=0$}, using Eq.~\eqref{csquared} for 
$C^2$, the anyon-anyon interaction matrix element reduces to
\begin{equation}\label{aaaa}
\langle 0|\hat{V}_{aa}|0\rangle=\frac{1}{5}\sqrt{\frac{\pi}{5}}.
\end{equation}

\subsection{Anyon-hole interaction}\label{SS:anyonhole}

The anyon-hole interaction takes the form
\begin{equation}
\hat{V}_{ah}=-\frac{1}{5}\sum_{j=1}^5\;\hat{V}_{jh}=
-\frac{1}{5}\sum_{j=1}^5\;\frac{1}{|\mathbf{r}_{jh}|},
\end{equation}
where
\begin{equation}
\mathbf{r}_{jh}=\boldsymbol{\xi}_j-\boldsymbol{\rho}+h\hat{\mathbf{z}}.
\end{equation}
Considering only $\hat{V}_{1h}$, we take the Fourier transform:
\begin{equation}
\hat{V}_{1h}\left(\boldsymbol{\rho},{\boldsymbol{\xi}}_1\right)=
-\frac{1}{5}\int\frac{d\mathbf{q}}{(2\pi)^2}V_{ah}(q)
\exp\left\{i\mathbf{q}\cdot\left({\boldsymbol{\xi}}_1-\boldsymbol{\rho}
\right)\right\},
\end{equation}
where
\begin{equation}
V_{ah}(q)=\frac{2\pi}{q}e^{-qh}.
\end{equation}
The matrix elements are then
\begin{multline}
\langle\left\{n\right\}|\hat{V}_{1h}|\left\{n'\right\}\rangle =
-\frac{1}{5}\int d\mathbf{R}
\int d\boldsymbol{\rho}\int\frac{d\mathbf{f}}{(2\pi)^2}
\int\frac{d\mathbf{q}}{(2\pi)^2}V_{ah}(q)
\int d\boldsymbol{\xi}_1 d\boldsymbol{\xi}_2 d\boldsymbol{\xi}_3 
d\boldsymbol{\xi}_4 d\boldsymbol{\xi}_5 \\
\times\bar{\Psi}_{{\{n\},\mathbf{k}}}\Psi_{\{n'\},\mathbf{k}}
\exp\left\{i\mathbf{f}\cdot
\left(\boldsymbol{\xi}_1+\boldsymbol{\xi}_2+\boldsymbol{\xi}_3
+\boldsymbol{\xi}_4+\boldsymbol{\xi}_5\right)\right\}
\exp\left\{i\mathbf{q}\cdot\left({\boldsymbol{\xi}}_1-\boldsymbol{\rho}
\right)\right\}.
\end{multline}
Integration over $\mathbf{R}$ and $\boldsymbol{\rho}$ gives
\mbox{$2\pi A\exp\left(-q^2/2-i\mathbf{d}\cdot\mathbf{q}\right)$}.
Then, following the procedure in Section \ref{SS:overlap} once again,
we obtain
\begin{multline}
\langle\left\{n\right\}|\hat{V}_{1h}|\left\{n'\right\}\rangle =
-\frac{C^2A(2\pi)^2}{5}\prod_{j=1}^5 i^{|n_j-n'_j|}
\int d\mathbf{f}\int d\mathbf{q}\;V_{ah}(q)
e^{-q^2/2-i\mathbf{d}\cdot\mathbf{q}} \\
\times\exp\left\{i\phi_{+}(n_1-n_1^{\prime})
+i\phi_{\mathbf{f}}\left[(L-L^{\prime})-(n_1-n_1^{\prime})
\right]\right\}
\mathcal{M}_{n_1 n'_1}(f_{+})\prod_{j=2}^5
\mathcal{M}_{n_j n^{\prime}_j}(f),
\end{multline}
where $\mathcal{M}_{mm'}$ are as defined in Eq.~\eqref{curlym}, 
\mbox{$\mathbf{f}_{+}=\mathbf{f}+\mathbf{q}$}, and $\phi_{+}$
is the phase of $\mathbf{f}_{+}$.
As before, \mbox{$L=n_1+n_2+n_3+n_4+n_5$}.

We now eliminate $\phi_{\mathbf{f}}$ by substituting 
\mbox{$\phi=\phi_{\mathbf{f}}-\phi_{\mathbf{q}}$} and
\mbox{$\psi=\phi_{+}-\phi_{\mathbf{f}}$}. Choosing the $x$-axis along 
$\mathbf{d}$, the integration over $\phi_{\mathbf{q}}$ is then
\begin{equation}
\int_0^{2\pi}d\phi_{\mathbf{q}}\;\exp\left\{i\phi_{\mathbf{q}}
(L-L^{\prime})-idq\cos\phi_{\mathbf{q}}\right\}
=2\pi{(-i)}^{|L-L^{\prime}|}J_{|L-L^{\prime}|}(dq),
\end{equation}
which gives for the matrix elements
\begin{multline}\label{ahole}
\langle\left\{n\right\}|\hat{V}_{1h}|\left\{n'\right\}\rangle =
-\frac{C^2(2\pi)^4A}{5}{(-i)}^{|L-L'|}\prod_{j=1}^5 i^{|n_j-n'_j|}
\int_0^{\infty}dq\:e^{-q^2/2-qh}J_{|L-L'|}(dq) \\
\times\int_0^{\infty}df\: f\int_0^{2\pi}d\phi\;
\exp\left\{i\phi(L-L^{\prime})+i\psi(n_1-n_1^{\prime})\right\}
\mathcal{M}_{n_1 n'_1}(f_{+})
\prod_{j=2}^5
\mathcal{M}_{n_j n^{\prime}_j}(f),
\end{multline}
where $\psi$ and $f_{+}$ can be expressed in terms of the variables of
integration as
\begin{equation}
e^{i\psi}=\left(f+qe^{-i\phi}\right)/f_{+},\qquad
f_{+}^2=f^2+q^2+2fq\cos\phi.
\end{equation}
For $d=0$, the Bessel function $J_{|L-L'|}(dq)$ entering the integrand
of Eq.~\eqref{ahole} is non-zero only if \mbox{$L=L^{\prime}$}.
This indicates that for the case of zero in-plane momentum \mbox{$(k=d=0)$},
the anyon-hole interaction matrix has the same block-diagonal
structure as the anyon-anyon and overlap matrices.
Therefore, states with different angular momentum $L_z$ decouple.

Eq.~\eqref{ahole} may be simplified further, and this yields the
following expression for the anyon-hole interaction matrix elements
for basis functions in terms of symmetric polynomials:
\begin{equation}\label{Qpoly}
\langle L,M|\hat{V}_{ah}|L',M'\rangle =
-\int_0^{\infty}dq\;\exp(-5q^2/2-qh)J_{|L-L'|}(dq)Q_{LM,L'M'}(q),
\end{equation}
where $Q_{LM,L'M'}(q)$ is a polynomial in $q$ (the lowest order
polynomial is \mbox{$Q_{00,00}=1$}). A similar expression to Eq.~\eqref{Qpoly} 
is obtained for a four-particle exciton in Ref.~\onlinecite{Portnoi1996}.
For \mbox{$h=0$}, Eq.~\eqref{Qpoly} further reduces to an
expression in terms of elementary functions.  For the case of \mbox{$k=d=0$} 
and \mbox{$L=L^{\prime}=0$}, the anyon-hole interaction matrix element is then
\begin{equation}\label{ahah}
\langle 0|\hat{V}_{ah}|0\rangle=-\sqrt{\frac{\pi}{10}}.
\end{equation}

\subsection{Exact results for $\mathbf{k}=0$, $L=0$}

All the above results are simplified significantly for the state
with \mbox{$\mathbf{k}=0$} and \mbox{$L=0$}. 
For such a state, the \mbox{$(N+1)$}-particle wavefunction 
\eqref{finalbasis} with \mbox{$\alpha=0$} reduces to the form
\begin{equation}
\Psi\left(\mathbf{R},\boldsymbol{\rho},\{\zeta_i\}\right)=
C_N\exp\left\{i\hat{\mathbf{z}}\cdot\left[\mathbf{R}\times\boldsymbol{\rho}
\right]/2-\rho^2/4\right\}\exp\left\{-\sum_{p=1}^N|\zeta_p|^2/4N\right\},
\end{equation} 
where the normalization constant $C_N$ can be determined from the integral
\vspace{3mm}
\begin{equation}
C_N^2\int d\mathbf{R}\int\frac{d\mathbf{f}}{{(2\pi)}^2}
\int d\boldsymbol{\rho}\;e^{-\rho^2/2}\int d\boldsymbol{\xi}_1
\cdots d\boldsymbol{\xi}_N
\exp\left\{\sum_{p=1}^N\left(
i\boldsymbol{\xi}_p\cdot\mathbf{f}-\xi_p^2/2N\right)\right\}
=\frac{C_N^2 A}{N^2}(2\pi N)^N=1.
\vspace{3mm}
\end{equation}

For the case of $\mathbf{k}=0$ and $L=0$, the interaction matrix elements
of a \mbox{$(N+1)$}-particle exciton are straightforward to evaluate.
The anyon-anyon matrix element is
\begin{align}\label{anyany}
\langle 0|\hat{V}_{aa}|0\rangle &=\frac{N(N-1)}{2A(2\pi N)^N}
\int d\mathbf{R}\int\frac{d\mathbf{f}}{(2\pi)^2}
\int\frac{d\mathbf{q}}{2\pi q}\int d\boldsymbol{\rho}
\;e^{-\rho^2/2}
\int d\boldsymbol{\xi}_1
\cdots d\boldsymbol{\xi}_N \notag \\
&\times\exp\left\{\sum_{p=1}^N\left(
i\boldsymbol{\xi}_p\cdot\mathbf{f}-\xi_p^2/2N\right)+i\mathbf{q}\cdot
\left(\boldsymbol{\xi}_1-\boldsymbol{\xi}_2\right)\right\}
=\frac{(N-1)}{4N}\sqrt{\frac{\pi}{N}}.
\end{align}
For non-zero inter-plane separation $h$ the anyon-hole
matrix element can be reduced to
\begin{align}\label{erfc}
\langle 0|\hat{V}_{ah}|0\rangle &=-\frac{N^2}{A(2\pi N)^N}
\int d\mathbf{R}\int\frac{d\mathbf{f}}{(2\pi)^2}
\int\frac{d\mathbf{q}}{2\pi q}\int d\boldsymbol{\rho}
\;e^{-\rho^2/2} \notag \\
&\times\int d\boldsymbol{\xi}_1
\cdots d\boldsymbol{\xi}_N\;\exp\left\{\sum_{p=1}^N\left(
i\boldsymbol{\xi}_p\cdot\mathbf{f}-\xi_p^2/2N\right)+i\mathbf{q}\cdot
\left(\boldsymbol{\xi}_1-\boldsymbol{\rho}\right)-qh \right\} \notag \\
&=-\int_0^{\infty}dq\;\exp (-Nq^2/2-qh)
=-\sqrt{\frac{\pi}{2N}}\:e^{h^2/2N}\mbox{erfc}\left( h/\sqrt{2N}\right),
\end{align}
where $\mbox{erfc}(x)$ is the complementary error function.
Using the asymptotic expansion of $\mbox{erfc}(x)$ it can be easily
seen from Eq.~\eqref{erfc} that 
\mbox{$\langle 0|\hat{V}_{ah}|0\rangle
\rightarrow -1/h$} as \mbox{$h\rightarrow\infty$}, as expected.

Eqs.~\eqref{anyany} and \eqref{erfc} also allow us to calculate the
critical inter-plane separation $h_c$ at which the \mbox{$\mathbf{k}=0$},
\mbox{$L=0$} state becomes unbound, i.e. when
\begin{equation}
\sqrt{\frac{\pi}{2N}}\;e^{x_c^2}\;\mbox{erfc}(x_c)=
\frac{(N-1)\sqrt{2}}{4N}\sqrt{\frac{\pi}{N}},
\end{equation}
where $x_c=h_c/\sqrt{2N}$. So, for \mbox{$N=3$} the critical separation 
\mbox{$h_c\approx 5.39\:l_H$}, for \mbox{$N=5$} we find that 
\mbox{$h_c\approx 5.59\:l_H$}, and for \mbox{$N\gg 1$} we have 
\mbox{$h_c\approx 1.32\sqrt{2N}\:l_H$}. 
Notably, these critical 
separations are well inside the region for which the AEM is applicable.
It should be emphasized that the state with \mbox{$L=0$} is not the ground
state for the anyon exciton at large separation $h$.
For example, for a four-particle exciton,\cite{Portnoi1996}
the ground states for large separation satisfy a superselection rule
\mbox{$L=3m$}, where $m$ is an integer, and when \mbox{$h\rightarrow\infty$} 
the ground state energy tends to its classical value, 
\mbox{$\varepsilon=-(2/3)^{3/2}/h$}.

Despite the non-applicability of our model to a real physical situation
at small layer separations \mbox{$(h<l_H)$}, the model remains soluble for
all values of $h$, including \mbox{$h=0$}. Moreover, it has been shown 
in Ref.~\onlinecite{Portnoi1996} that
the ground state of the four-particle problem \mbox{$(N=3)$} at \mbox{$h=0$} 
is the state with \mbox{$\mathbf{k}=0$} and \mbox{$L=0$}. 
We expect the same to be true for \mbox{$N\geqslant 5$}, since the 
anyon-hole attraction will always overcome the anyon-anyon repulsion 
at small inter-particle separations. 
It can be shown (in a similar way to that in
Ref.~\onlinecite{Portnoi1996}) that the smallest average inter-particle 
separation corresponds to the \mbox{$L=0$} case. Therefore, the case of
\mbox{$\mathbf{k}=0$}, \mbox{$L=0$} is of special interest since it 
predicts the anyon exciton binding energies in the limit 
\mbox{$h\rightarrow 0$}.

We are now in a position to write down a general expression for the
binding energy of a \mbox{$(N+1)$}-particle exciton at zero 2DEG-hole
separation \mbox{$(h=0)$}:
\begin{equation}\label{bind}
E_b=-\left(\langle 0|\hat{V}_{ah}|0\rangle
+\langle 0|\hat{V}_{aa}|0\rangle\right)
=\left[1-\frac{(N-1)}{2\sqrt{2}N}\right]
\sqrt{\frac{\pi}{2N}}.
\end{equation}
Eq.~\eqref{bind} has been written in this particular form to emphasize
the key result that for any value of $N$ there always exists at least
one bound state of a neutral \mbox{$(N+1)$}-particle anyon exciton at 
\mbox{$h=0$}.  
For \mbox{$N=1$}, Eq.~\eqref{bind} yields the value \mbox{$\sqrt{\pi/2}$},
which corresponds to that obtained in Ref.~\onlinecite{Lerner1980} for a
standard diamagnetic exciton. For a four-particle exciton \mbox{$(N=3)$},
\mbox{$E_b=\left(1-\sqrt{2}/6\right)\sqrt{\pi/6}$}, 
which agrees with the result of Ref.~\onlinecite{Portnoi1996}. 
Finally, for \mbox{$N=5$} we have a binding energy of
\mbox{$(1-\sqrt{2}/5)\sqrt{\pi/10}$}, which can also be obtained from
Eqs.~\eqref{aaaa} and \eqref{ahah}.

\section{Conclusions}

The anyon exciton model has been generalized to the case of an arbitrary
number of anyons and several important mathematical results have been 
obtained. 
Starting from the Hamiltonian for a non-interacting system of $N$ anyons
and a valence hole in a quantizing magnetic field, we have obtained a 
complete set of exciton basis functions.
These functions have been fully classified using a result from the
theory of partitions.
We have derived expressions for the overlap and interaction matrix 
elements for a six-particle system, which describes an exciton against
the background of an IQL with filling factor \mbox{$\nu=2/5$}.
In the particular case of \mbox{$\mathbf{k}=0$} and \mbox{$L=0$}, we have 
found an expression for the binding energy of a 
\mbox{$(N+1)$}-particle exciton, which agrees with known results for a 
standard diamagnetic exciton and a four-particle anyon exciton.
We have also shown that the \mbox{$(N+1)$}-particle exciton remains bound 
for 2DEG-hole separations exceeding several magnetic lengths, when the 
anyon exciton model becomes applicable to real physical systems.


\bibliographystyle{apsrev}



\end{document}